\documentstyle[12pt,epsfig]{article}
\begin{document}

\def\ii{\'{\char'20}}
\def\r{\rightarrow}
\def\err{\end{array}}
\def\bea{\begin{eqnarray}}
\def\eea{\end{eqnarray}}
\newcommand{\beq}{\begin{equation}}
\newcommand{\eeq}{\end{equation}}
\newcommand{\nn}{\nonumber}

\begin{titlepage}

\title{
{\Large \bf  Study of Wilson loop functionals in 2D Yang-Mills 
theories}\thanks{Contribution to the Proceedings of the XIIth Workshop 
on High Energy Physics and Quantum Field Theory (Samara, Russia, 
September 4-10, 1997).}}

\author{J.M. Aroca\thanks{E-mail address: jomaroca@mat.upc.es} \\
Departament de Matem\`atica Aplicada i Telem\`atica \\ 
Universitat Polit\`ecnica de Catalunya \\ 
C./ Jordi Girona 1 i 3, Mod. C-3, Campus Nord \\08034 Barcelona, 
Spain \\
and  \\
Yu.A. Kubyshin 
\thanks{E-mail address: kubyshin@theory.npi.msu.su} \\
Institute for Nuclear Physics \\
Moscow State University \\ 
117899 Moscow, Russia}

\date{January, 1998}

\maketitle

\begin{abstract}
The derivation of the explicit formula for the vacuum expectation
value of the Wilson loop functional for an arbitrary gauge
group on an arbitrary orientable two-dimensional manifold is
considered both in the continuum case and on the lattice. 
A contribution to this quantity, coming from the space of 
invariant connections, is also analyzed and is shown 
to be similar to the contribution of monopoles. 
\end{abstract}

\end{titlepage}

\section{Introduction}

Yang-Mills theory in two dimensions has been an object of
intensive studies during almost two decades. It is known that
the classical theory is trivial and the quantum theory has no 
propagating degrees of freedom. However, this theory possesses
very many interesting properties when formulated on
topologically nontrivial spaces \cite{Witten} and also in the
large $N$ limit ($N$ is related to the rank of the gauge group
$G=SU(N)$) \cite{DK}. A 2D Yang-Mills theory is almost
topological in the sense that, for example, for compact
spacetime manifolds of area (two-dimensional volume) $V$ 
its partition function
depends only on $e^{2} V$, where $e$ is the gauge coupling
constant. 
Also, as we will see, the vacuum expectation value of 
a Wilson loop depends only on the area of the 
region surrounded by the loop and is independent of the points 
where the loop is located.   
This is a manifestation of the invariance under the
area preserving diffeomorphisms. It is believed that 2D
Yang-Mills theories share some of the important qualitative
features of 4D ones, in particular the area law behaviour. 
Recently there has been a revival of interest in
studying 2D Yang-Mills theories because it was shown that in the
strong coupling limit they can be related to 
lower-dimensional strings \cite{Gross} and because of a 
special role such theories play in 
M-theory \cite{YM-M}. Rich mathematical structures appearing in
two-dimensional gauge theories were studied in a number of
papers \cite{Witten}, \cite{AB} - \cite{BT1} (see lecture \cite{Moore}
for a review and references). The partition function and the
vacuum expectation values of Wilson loop functionals were
calculated by many authors with various techniques \cite{W-calc} - 
\cite{ALMMT}. Physical aspects of these theories, in particular 
the analysis of Polyakov loops and the $\theta$-vacua were
discussed in Refs. \cite{Grig}. 

In the present contribution we analyze vacuum expectation values
of the Wilson loop functionals. As a (Euclidean) space-time 
a smooth two-dimensional orientable Riemannian manifold $M$
will be considered. The gauge group 
$G$ is supposed to be a compact Lie
group. The gauge potential $A_{\mu}$ can be characterized by the 
1-form $A=A_{\mu} dx^{\mu}$ on $M$ associated with the connection 
form on the principal fibre bundle $P(M,G)$ \cite{KN} 
in a standard way via local sections (see Refs. \cite{gauge-geom} 
for geometrical desciption of gauge theories). 
We denote by ${\cal A}$ the space of smooth connections. 

Let us fix a point $x_{0}$ in $M$ and consider based loops
defined in a standard way as continuous mappings of the unit
interval $I=[0,1]$ into $M$:
\[ 
\gamma: I \r M, \; \; \; s  \rightarrow \gamma(s) \in M, 
\; \; \; s \in I
\]
with $\gamma(0)=\gamma(1)=x_{0}$. For a given $A \in {\cal A}$
we associate with each loop $\gamma$ the element of $G$ called 
holonomy 
\beq
H_{\gamma}(A) = {\cal P} \exp \left( ie\oint_{\gamma} A \right),
          \label{H-def}
\eeq 
where ${\cal P}$ means the path ordering. We will be 
interested in traced holonomies 
\beq
  T_{\gamma}(A) \equiv \frac{1}{d_{R}} Tr R \left(H_{\gamma}(A) \right) = 
  \frac{1}{d_{R}} \chi_R (H_{\gamma}(A)),
   \label{trh}
\eeq
called also Wilson loop variables \cite{Wi}, where $R$ is an
irreducible representation of $G$, $d_{R}$ is its dimension and
$\chi_R$ is its character. Let us denote by ${\cal T}$ the
group of local gauge transformations, i.e. the group of smooth
vertical automorphisms of $P$. It is known that $T_{\gamma}$
suffice to separate points in ${\cal A}/{\cal T}$ representing
classes of gauge equivalent configurations and, therefore, enable
to reconstruct all smooth gauge connections up to gauge
equivalence \cite{Giles}. Due to this property the Wilson loop
functionals (\ref{trh}) form a natural set of gauge invariant
functions of connections in the classical theory.

We will study the vacuum expectation
value $<T_{\gamma}>$ of the Wilson loop functional given by 
\beq
< T_{\gamma} > = \frac{1}{Z(0)} Z(\gamma), \; \; \; 
  Z(\gamma) =  \int {\cal D}A e^{-{\cal S}} T_{\gamma}(A),   
 \label{Zg-def}
\eeq
where ${\cal S}$ is the Yang-Mills action.
In Sect. 2 we will discuss as an example calculation of 
(\ref{Zg-def}) 
in the abelian case for arbitrary $M$. In Sect. 3 $<T_{\gamma}>$ 
will be calculated on the lattice for a general gauge group.   
In Sect. 4 we will study a special class of connections, 
namely invariant connections, and their contribution to $<T_{\gamma}>$ 
in the case $M=S^{2}$ and $G=SU(2)$. In Sect. 5  
this contribution will be compared with the complete vacuum expectation 
value (\ref{Zg-def}).  

\section{Continuum case, $G=U(1)$}

The action ${\cal S}$ in (\ref{Zg-def}) is given by the standard
expression 
\[
{\cal S} = \frac{1}{4}\int_{M} Tr \left( F \wedge \ *F \right)= 
 \frac{1}{8}\int_{M} Tr \left( F_{\mu \nu} F_{\mu \nu} \right)
 \sqrt{ \det g_{\mu \nu}} d^{2}x.
\]
In general the space of connections ${\cal A}$ consists of a number of
components, or sectors ${\cal A}_{\alpha}$, labelled by elements 
$\alpha$ of an index set ${\cal B}$, 
$\alpha \in {\cal B}$. 
The functional integral is represented as a sum
over the elements of ${\cal B}$, each term of the sum
being the functional integral over the connections in ${\cal
A}_{\alpha}$. This feature has an analog in quantum mechanics:
in the case of multiply connected space $M$ evaluation of the functional
integral includes summation over the 
connected components of the space of paths in $M$ and
integration over the paths within each component \cite{QM-analog}. 

The set ${\cal B}$ of connected 
components of ${\cal A}$ is in 1-1 correspondence with the space of 
non-equivalent principal $G$-bundles $P(M,G)$ over the manifold $M$. 
Let us denote this space as ${\cal B}_{G}(M)$, ${\cal B} \cong
{\cal B}_{G}(M)$. The problem of classification of fibre bundles
was considered in a number of books and articles. 
Following closely the lectures by Avis and Isham \cite{AI} we
obtained that in the case when $M$ is a two-dimensional
manifold (actually, even a CW-complex) ${\cal B}_{G}(M)$ is equal to 
\[
   {\cal B}_{G}(M) \cong H^{2} (M,\pi_{1}(G)),    
\]
the second cohomology group of $M$ with coefficients given by 
elements of the first homotopy group of the gauge group $G$ 
(see details in \cite{AK}).
An equivalent classification of principal fibre bundles over a
two-dimensional surface in terms of elements of the group
$\Gamma$, specifying the global structure of $G$ through the
relation $G = \tilde{G}/\Gamma$, where $\tilde{G}$ is the
universal covering group, was given in the second article of
Ref. \cite{Witten}. 
For completeness, we present here a list of the first homotopy groups 
$\pi_{1}(G)$ for  some Lie groups which are of interest in gauge theories: 
$\pi_{1}(SU(n)) = \pi_{1}(Sp(n)) = 0$, $\pi_{1}(SO(n)) = Z_{2}$
($n = 3$ and $n \geq 5$), $\pi_{1}(U(n)) = Z$.

Let us consider the abelian case.  $U(1)$-bundles are classified
by elements of the second cohomology group $H^{2}(M,Z)$. The
class $c_{1}(P) \in H^{2}(M,Z)$, corresponding to the bundle $P$,
is known as the first (integer) Chern class  
\cite{KN}, \cite{Span}. For a closed orientable 
2-dimensional manifold $M$ $H^{2}(M,Z) \cong Z$. 
For the bundle $P_{n}$, characterized by the label $n \in {\cal B} \cong
Z$, the integer cohomology class can be represented by 
$e F^{(n)} / (2 \pi)$ with $F^{(n)}$ being the curvature 2-form 
defined locally through $F^{(n)} = dA^{(n)}$, where $A^{(n)}$ is the 
gauge 1-form given by a connection on $P_{n}$. The integral 
\[
    \frac{e}{2 \pi} \int_{M}  F^{(n)} = n, 
\]
does not depend on the choice of the connection and gives the
1st Chern number, also called the topological charge.

From the discussion above we conclude that 
\beq
Z(\gamma) = \sum_{n=-\infty}^{\infty} \int {\cal D} A^{(n)} 
   e^{-{\cal S}(A^{(n)})} T_{\gamma}(A^{(n)}) = \sum_{n=-\infty}^{\infty}
    Z^{(n)}(\gamma),          \label{Z-sum}
\eeq
where in each term $Z^{(n)}$ the integration goes over the gauge
potentials with the Chern number equal to $n$. We
represent $A^{(n)}$ as $A^{(n)} = \tilde{A}^{(n)} + a$, where
$\tilde{A}^{(n)}$ is a potential with
\beq
     \int_{M} \tilde{F}^{(n)} = \frac{2\pi n}{e},  \label{F-int}
\eeq
$\tilde{F}^{(n)} = d \tilde{A}^{(n)}$, and $a$ is a 1-form with
zero Chern number.  As $\tilde{A}^{(n)}$ we take instanton 
configurations, i.e. solutions of the equation of motion. In the 
literature they are often referred to as monopoles, and we will 
follow this 
terminology in the present paper. 
Then $Z^{(n)}(\gamma)$ in (\ref{Z-sum}) becomes 
\bea
Z^{(n)}(\gamma) & = & e^{-{\cal S}(\tilde{A}^{(n)})} T_{\gamma} 
(\tilde{A}^{(n)}) Z_{0}(\gamma), 
     \label{Zn}  \\
Z_{0}(\gamma) & = & \int_{{\cal A}^{(0)}} {\cal D}a e^{-{\cal S}(a)} 
   T_{\gamma}(a),    \label{Z0}
\eea
and the functional integral in $Z_{0}(\gamma)$ goes over the
connections in the trivial bundle $P_{0}$ with zero topological
charge. 

It turns out that for the evaluation of (\ref{Zn}) we 
do not need explicit expressions for the monopole solutions 
$\tilde{A}^{(n)}$. Indeed, it is easy to show using Eq.
(\ref{F-int}) that 
\[
{\cal S}(\tilde{A}^{(n)}) = \frac{2\pi^{2} n^{2}}{Ve^{2}}.    
\]
If $\gamma = \partial \sigma$, where $\sigma$ is the interior of
$\gamma$ and $\tilde{A}^{(n)}$ is regular in $\sigma$, 
then by using Stoke's theorem we obtain that 
\[
\oint_{\gamma} \tilde{A}^{(n)}  = 
\oint_{\partial \sigma} \tilde{A}^{(n)} = 
\int_{\sigma} d\tilde{A}^{(n)} =  
 \int_{\sigma} \tilde{F}^{(n)} = 
\frac{2\pi n}{Ve} S,  
\]
where we denoted the area of the surface $\sigma$ by $S$. 
Then the contribution of the monopoles to the full function 
(\ref{Z-sum}) is equal to  
\bea
 & & Z_{mon}(\gamma) := \sum_{n=-\infty}^{\infty} e^{-{\cal S}
(\tilde{A}^{(n)})} T_{\gamma} (\tilde{A}^{(n)})   \nonumber \\
  & &  =  \sum_{n=-\infty}^{\infty} \exp 
  \left( -\frac{2\pi^{2}n^{2}}{Ve^{2}} + 
  i \frac{2\pi n}{V}\nu S \right) = 
 e \sqrt{\frac{V}{2\pi}} \sum_{l=-\infty}^{\infty} 
  \exp \left[ - \frac{Ve^{2}}{2} \left( \frac{S}{V}\nu + l 
 \right)^{2} \right].   \label{Zmon}
\eea
The last equality was obtained using the Poisson summation
formula. We have taken the Wilson loop variable in an
arbitrary irreducible representation so that 
$T_{\gamma}(A) = \exp \{ i\nu e
\oint_{\gamma} A\}$, $\nu$ being an integer. 
 
Now let us turn to the calculation of the contribution 
$Z_{0}(\gamma)$ given by the functional integral Eq. (\ref{Z0}).
The power of the exponent in $T_{\gamma}(A)$ can be understood
as a functional $J_{\gamma}$ acting on $A$ and written
as 
\[
J_{\gamma}[A] \equiv e \oint_{\gamma} A = 
\int_{M} dv A_{\mu}(x) J_{\gamma}^{\mu}(x).  
\] 
Then the integral (\ref{Z0}) becomes gaussian and can be easily
calculated. However, the components $J_{\gamma}^{\mu}$ are not
smooth functions, 
\[
  J_{\gamma}^{\mu}(y) = e\oint_{\gamma} dx^{\mu} \frac{1}
{\sqrt{\det g_{\mu \nu}}} \delta (y-x),
\]
so that $J_{\gamma}[A]$ should be treated as a singular form, or
weak form, \cite{Fels} for the accurate calculation of the
contribution of fluctuations around the monopoles. The result is 
\[
Z_{0}(\gamma) = {\cal N} e^{-\frac{e^{2}}{2}\nu^2
\frac{S(V-S)}{V}}.  
\]

Combining this with the contribution (\ref{Zmon}) of monopoles
we obtain the final expression for the 
expectation value of the Wilson loop variable for a 
homologically trivial loop $\gamma$ in the abelian case: 
\beq
<T_{\gamma}> = \frac{
  \sum_{l=-\infty}^{\infty} \exp \left[ - \frac{e^{2}}{2} V 
  \left( \frac{S}{V}\nu + l \right)^{2} - \frac{e^{2}}{2}\nu^2 
  \frac{S(V-S)}{V} \right]}{\sum_{l=-\infty}^{\infty} 
  \exp \left[ - \frac{e^{2}}{2} V l^{2} \right]}. 
  \label{Tg}
\eeq
Expression (\ref{Tg})
depends only on the total area $V$ of the compact
two-dimensional manifold $M$ and the area $S$ of $\sigma$
surrounded by $\gamma$. It is invariant under the transformation
$S \rightarrow (V - S)$, as it must be, because any of the two regions
of $M$, in which the loop $\gamma$ divides it, can be considered
as the interior of $\gamma$. 

\section{Two-dimensional gauge theories on the lattice}

Here we consider the case of arbitrary gauge group $G$ and arbitrary
compact orientable two-dimensional manifold $M$. 
We assume that $\gamma$ divides $M$ into two regions
$\sigma_1, \sigma_2$ of genera $r_1, r_2$ and areas $S_1, S_2$
respectively. The genus of $M$ is $r=r_1+r_2$ and its area
$V=S_1+S_2$. 

To perform the calculation we introduce a connected graph, or
lattice, $\Lambda$ on
the manifold $M$ which is a union of finite sets whose elements
are 0-cells $s$ (sites), 1-cells $l$ (links) and 2-cells
$p$ (plaquettes). We consider only such lattices that the closed path
$\gamma \subset \Lambda$, that is $\gamma$ is a sequence of
links of $\Lambda$. We denote the area
of the plaquette $p$ as $|p|$. The ``lattice spacing" $a$
is defined as the minimum distance such that every plaquette is
contained in a circle of the radius $a$. As a
system of loops let us choose  $\gamma_{p} = \partial p$, i.e.
loops which are boundaries of the plaquettes, and a set of 
generators $a_i, b_i$ of the fundamental group of $M$. 
The index $i=1,\ldots ,g$ labels the handles of $M$.
Any loop on the lattice can be obtained from the system 
$\{\gamma_{p},a_i,b_i\}$.
The loops $a_{i}$, $b_{i}$ allow us to be in any homotopy class while
the loops $\gamma_{p}$ generate continuum deformations. This system
is not independent but subjected to one relation:
\beq
      \prod_{p \in \Lambda} \gamma_{p} = 
     \prod_{i=1}^g a_ib_ia_i^{-1}b_i^{-1}    \label{loop-rel}
\eeq
from which one can construct an independent system of loops
$\beta_{j}$ $(j = 1, 2, \ldots, n = \dim \pi_{1}(\Lambda))$.
The product in (\ref{loop-rel}) has to be taken in a certain order
which can be shown to exist.  
Also we would like to note that for the rigorous treatment
independent hoops, that are classes of holonomically equivalent
loops, should be considered. We will skip this distinction here.

Using results of \cite{AL} (see also \cite{ALMMT}) one can give a 
well-defined meaning to the heuristic measure ${\cal D}A $ in Eq.
(\ref{Zg-def}) as a measure constructed out of copies of the 
Haar measure on the
group $G$. This is possible if the action can be written as a
cylindrical function on the space ${\cal A} / {\cal T}$.
The standard Yang-Mills action 
${\cal S}(A)$ is not well defined on such space, hence 
one has to use some regularized action ${\cal S}_{\Lambda}(A)$ 
on the lattice $\Lambda$, calculate $Z_{\Lambda}(\gamma)$, which
is a lattice analog of (\ref{Zg-def}), and then take the limit 
$Z(\gamma) = \lim_{a \rightarrow 0} Z_{\Lambda}(\gamma)$. 
 
In this article we consider a class of actions such that 
\[
   e^{-{\cal S}_{\Lambda}(A)} = \prod _{p \in \Lambda}  
  e^{-{\cal S}_{1}(H_{\gamma_{p}}(A))},  
\]
where the holonomy $H_{\gamma}(A)$ is defined by Eq.
(\ref{H-def}). 
The function ${\cal S}_{1}$ is a real function over $G$
which satisfies the following conditions: 
1) ${\cal S}_{1}(g) = {\cal S}_{1}(g^{-1})$; 
2) it has the absolute minimum on the identity element;
3) $\lim_{a \rightarrow 0}{\cal S}_{1}(H_{p}(A))|p|^{-1} = 
Tr( F_{\mu \nu}(x))^{2}/2$.
An important example is given by the Wilson action: 
${\cal S}_{1}(g) = 1 - Re \chi_F (g)/d_{F}$, where $F$ stands for the 
fundamental representation.

With such actions the integrand in (\ref{Zg-def}) is 
a cylindrical function. Let us write the action in terms of the Fourier
coefficients:
\beq
\mu_{p}(g) \equiv e^{-{\cal S}_1(g)} = \sum_R d_R \chi_{R}(g)
B_{R}(p),    \label{mu-p}
\eeq
%where the coefficients $B_{R} (p)$ are
%\beq
%B_{R} (p)=\int dg e^{-{\cal S}_1(g)}\chi_R^{*}(g).   \label{B-R}
%\eeq
where the sum goes over all irreducible representations of $G$. 
Any action can be specified through the coefficients $B_{R}$.
A relevant case is the heat-kernel action given by
$B_{R} (p) = (\tilde{B}_{R})^{|p|}$,
$\tilde{B}_{R} = \exp (- e^{2}c_2(R)/2)$,
where $c_{2}(R)$ is the value of the second Casimir operator. 
When $G=U(1)$ the heat-kernel action is called Villain action. 

Let us denote by $g_{a_{i}}$ and $g_{b_{i}}$ the holonomies 
corresponding to the generators $a_{i}$ and $b_{i}$ of the
homotopy group and by $g_{p} \equiv g_{\gamma_{p}}$ the holonomies 
corresponding to the loops $\gamma_{p}$. 
The functional integral $Z_{\Lambda} (\gamma)$ is equal to 
\beq
 Z_{\Lambda}(\gamma) = 
	\int_{G^{n+1}} (\prod _{i} dg_{a_i}dg_{b_i})
	\left( \prod _{p \in \Lambda} 
	dg_{p} \mu_{p}(g_{p})\right)\Delta [\{g_q\}]
	\frac{1}{d_R} \chi_R \left(\prod_{q
      \in \gamma} g_{q} \right).   \label{Zp-def}
\eeq
Here $dg$ is the Haar measure on $G$ and
$\Delta [\{g_q\}]$ is a factor which imposes the relation
between the group variables implied by (\ref{loop-rel}): 
\[
\Delta [\{g_q\}]=
\delta_G (\prod_{p\in \Lambda} g_{p} , 
\prod_i g_{a_i}g_{b_i}g_{a_i}^{-1}g_{b_i}^{-1} ), 
\]
where $\delta_G(g,h)$ is the Dirac distribution on the group $G$. 
A specific decomposition of $\gamma$, which
appeares in the argument of the character $\chi_{F}$ in
(\ref{Zp-def}), is given in terms of
the loops located in one of the regions, say, $\sigma_1$,
namely  
\[
\gamma =\prod_p^{(1)}\gamma_p 
[\prod_i^{(1)}a_ib_ia_i^{-1}b_i^{-1}]^{-1}
\]
where the superscript $(1)$ means the restriction to $\sigma_1$. 

Using Eq. (\ref{mu-p}) and interchanging 
the summation in (\ref{mu-p}) with the product over the
plaquettes in (\ref{Zp-def}) we obtain that 
\bea
Z_{\Lambda}(\gamma )  & = & 
 \sum_{\{R_p\}}(\prod_{p \in \Lambda} B_{R_p}(p))
\int_{G^{n+1}} (\prod _{i} dg_{a_i}dg_{b_i})
\left( \prod _{p \in \Lambda} 
	dg_{p}\right) (\prod _{p \in \Lambda} d_{R_p} 
	\chi_{R_p}(g_p))      \nonumber \\
   & \times &   \Delta [\{g_q\}] \frac{1}{d_R} \chi_R 
     \left(\prod_{q \in \gamma} g_{q} \right).   \nonumber 
\eea
Here the first sum goes over all possible ``colorations'' of the
surface, i.e. over all possible configurations of the
irreducible representations, where each configuration $\{ R_{p}
\}$ is a set of irreducuble representations assigned to
each plaquette. 

To perform the $g_p$ integration we use a procedure of
``lattice reduction". If plaquettes $p_1$ and $p_2$ share
a link that does not belong to $\gamma$ (then, they are
in the same $\sigma_i$ component) we change the group
variable $g_{p_1}\rightarrow g_{p_1}g_{p_2}^{-1}$. 
The measure of integration is invariant under such changes, 
$g_{p_2}$ disappears in $\Delta$ and $\chi_F$, and we can
integrate over $g_{2}$ using the properties of invariant
group integration.
The result is non-zero if both plaquettes are in the same
representation, in which case the common link is ``erased" and
the two plaquettes merge into one plaquette while the form of the
partition function remains the same. Such reductions can be done
until we arrive to a lattice consisting only of two plaquettes,
$p_1$ and $p_2$ with representations $R_1$ and $R_2$ respectively.
Integration over the variables of the homotopically non-trivial
loops gives rise to the factors $d_{R_{1}}^{1-2r_{1}}$ and
$d_{R_{2}}^{1-2r_{2}}$. Finally we obtain that 
\[
Z_\Lambda(\gamma )  = 
 \sum_{R_1, R_2}(\prod_{p \in \sigma_1} B_{R_1}(p))
                 (\prod_{p \in \sigma_2} B_{R_2}(p))
	d_{R_1}^{1-2r_1}d_{R_2}^{1-2r_2}\{R_1,R,R_2\},	
\]
where 
\[
\{R_1,R,R_2\} = \int_{G} dg \chi_{R_1}(g) 
   \chi_{R_2}(g^{-1}) \chi_{R}(g) 
\] 
is the number of times the representation
$R_2$ is contained in $R_1\otimes R$.

To obtain the continuum limit one should, in general,
compute the coefficients $B_{R}(p)$ and take the 
limit $|p| \rightarrow 0$. This is the case if we work
with the Wilson action. If instead we use the heat-kernel action,
$Z_{\Lambda} (\gamma)$ can be written in the form 
\[
Z_{\Lambda} (\gamma)  = 
 \sum_{R_1, R_2} \tilde{B}_{R_1}^{S_1}
                 \tilde{B}_{R_2}^{S_2}
	d_{R_1}^{1-2r_1}d_{R_2}^{1-2r_2}\{R_1,R,R_2\}
\]
which is lattice independent, so it gives already the continuum
limit. The final expression for the continuum limit of the
vacuum expectation value of
the Wilson loop functional is 
\beq
<T_\gamma >  = 
\frac{
 \sum_{R_1, R_2} \tilde{B}_{R_1}^{S_1}
                 \tilde{B}_{R_2}^{S_2}
	d_{R_1}^{1-2r_1}d_{R_2}^{1-2r_2}\{R_1,R,R_2\} }
{ d_{R}\sum_{R_3}\tilde{B}_{R_3}^{V}d_{R_3}^{2-2r}}.    \label{Tfin}
\eeq

For abelian groups $d_{R}=1$. In this case the topology of 
the surface is irrelevant. 
If $G=U(1)$ the irreducible
representations are labelled by integers $n\in Z$. Since 
$B_n =e^{-e^2n^2/2}$, taking $R=\nu$ we get
\[
<T_{\gamma} >=\frac{
\sum_{n=-\infty}^\infty B_n^{S_1}B_{n+\nu}^{S_2}}
{\sum_{n=-\infty}^\infty B_n^V}  
  =\frac{\sum_{n=-\infty}^\infty e^{-S_1e^2n^2/2} 
  e^{-S_2e^2(n+\nu)^2/2}} {\sum_{n=-\infty}^\infty 
  e^{-Ve^2n^2/2}}.    
\]
This coincides with (\ref{Tg}) after the identifications  
$S_{1}=S$ and $S_{2}=V-S$. 

We would like to make a few comments. The result (\ref{Tfin}) 
is invariant under the transformation $S \leftrightarrow V-S$. 
Although we have considered only compact surfaces,
the same techniques can be applied in the non-compact case.
The important difference is that the infinite component at one
side of $\gamma$ appears to be in the trivial representation.
This case can also be analized as the limit $V\rightarrow \infty$
of a compact case. Performing it in (\ref{Tfin}) we see
that the Wilson loop expectation value for $R^{2}$ 
or any non-compact
surface has the typical area law behaviour 
$<T_{\gamma}> \propto \exp \left(-e^{2}c_{2}(R) S/2 \right)$, 
where $S$ is the area of the compact part enclosed by $\gamma$.
For finite $V$ similar behaviour is obtained in
the strong coupling limit $e^2 V \gg 1$ 
in which case only the region with the smaller area 
contributes. In the weak coupling limit $e^2 V \ll 1$ for 
the abelian case 
\[
<T_{\gamma }>\sim e^{-e^2(S-S^2/V)\nu^2/2}.
\]

\section{Wilson loop variables for invariant connections.}

In Sect. 2, while calculating the vacuum expectation value of
the Wilson loop functional 
in the abelian case, we analyzed the contribution of the monopoles. 
In the present section we 
will study this quantity for a special class of 
connections, called invariant connections, for $M=S^{2}$ and $G=SU(2)$. 
The result of the calculation will be compared with the exact 
formula derived above. This will allow 
us to understand how much of the information is captured by the 
invariant connections. Our interest
in this class of connections is motivated by the fact that the 
analogous calculation can be 
carried out in Yang-Mills theory in any dimension provided, 
of course, that non-trivial invariant connections exist. 

Let us first discuss the definition of invariant connection  
and explain how they can be constructed. After that a  
concrete example with $M=S^{2}$ and $G=SU(2)$ will be considered. 
Assume that there is a group $K$ which acts on the
space-time $M$ and its action can be lifted to the fibre bundle
$P(M,G)$. We will call $K$ 
symmetry group. A connection in $P$ is said to be invariant with
respect to transformations $L_{k}$ of $K$ if its connection
1-form $w$ satisfies $L_{k}^{*} w = w$
for all $k \in K$. 
This class of connections includes many known monopole and instanton 
configurations and they were intensively
used for the coset space dimensional reduction of gauge theories
(see \cite{Manton}).
Gauge potentials corresponding to invariant
connections are called symmetric potentials and were introduced
in \cite{Schwarz}. The condition of invariance implies that for any
$k \in K$ there exists a gauge tranformation $g_{k}(x) \in G$
such that the gauge potential $A_{\mu}$, corresponding to the
invariant connection form $w$, has the property 
\beq
    (O_{k} A)_{\mu} = g_{k}(x)^{-1} A_{\mu}(x) g_{k}(x) + 
\frac{1}{ie}g_{k}(x)^{-1} \partial_{\mu} g_{k}(x),   \label{sympot}
\eeq
where the l.h.s. is the standard
change of the field under the space-time transformation $O_{k}$
on $M$ (which is the projection of $L_{k}$ to $M$). This
formula tells that a symmetric potential is invariant under
transformations of $K$ up to a gauge transformation.

Here we consider the case when $K$ acts transitively on $M$ 
(see Ref.
\cite{KN} for general theory and Ref. \cite{KMRV} for review). 
Then $M$ is a coset space $K/H$, where $H$ is a subgroup
of $K$, called the isotropy group, and $K$ acts on $K/H$ in the
canonical way. The 1-form $A = A_{\mu} dx^{\mu}$, which
describes a symmetric potential satisfying (\ref{sympot}) 
and is a pull - back of an
invariant form with respect to a (local) section  
of $P(M,G)$, can be constructed in the following way. 
Let ${\cal G}$, ${\cal K}$ and ${\cal H}$ be the Lie algebras 
of the groups $G$, $K$ and $H$ respectively. 
If $H$ is a closed compact subgroup of $K$, the case we 
have in mind, then there exists the decomposition
\beq
    {\cal K} = {\cal H} + {\cal M}     \label{K-decomp}
\eeq
with $ad ({\cal H}) {\cal M} = {\cal M}$, where $ad$ denotes
the adjoint action of the algebra on itself, i.e. $ad (h) (u) =
[h,u]$, $h,u \in {\cal K}$. Let $\theta$ be the canonical
left-invariant 1-form on the Lie group $K$ with values in ${\cal
K}$, and $\bar{\theta}$ is its pull-back to $M=K/H$. It can be
calculated as $\bar{\theta} = k(x)^{-1} dk(x)$, where $k(x) \in
K$ is a local representative of the class $x \in K/H$.  We
decompose the 1-form $\bar{\theta}$ into the ${\cal H}$- and
${\cal M}$-components in accordance with (\ref{K-decomp}):
$\bar{\theta} = \bar{\theta}_{\cal H} +
\bar{\theta}_{\cal M}$. It can be shown that 
the invariant connections are given by 
\beq
        A = \frac{1}{ie} \left( \tau (\bar{\theta}_{\cal H}) + 
        \phi (\bar{\theta}_{\cal M}) \right),   \label{A-inv}
\eeq
where $\tau$ is the homomorphism of algebras induced by a group
homomorphism $H \rightarrow G$ and $\phi$ is a mapping $\phi :
{\cal M} \rightarrow {\cal G}$ satisfying the equivariant
condition 
\beq
      \phi \left( ad (h) u \right) = ad (\tau (h)) \phi(u), 
    \; \; h \in {\cal H}, \; \; u \in {\cal M}.    \label{equiv1}
\eeq
This condition can be viewed as the intertwining condition
between representations of ${\cal H}$ in ${\cal M}$ and ${\cal
G}$. 
An effective technique for solving constraint (\ref{equiv1})
was developed in Refs. \cite{VK} (see also \cite{KMRV}). 

As a concrete example we will consider the Yang-Mills theory on
the two-dimensional sphere $S^{2}$ with the gauge group
$G=SU(2)$. The sphere is realized as a
coset space $S^{2} = SU(2)/U(1)$. Let us construct first the
1-forms $\bar{\theta}_{\cal H}$ and $\bar{\theta}_{\cal M}$
which appear in Eq. (\ref{A-inv}). As usual, we cover the
manifold $M=S^{2}$ with two charts $U_{1}$ and $U_{2}$, the
northern and southern hemispheres, so that $U_{1} \cup U_{2} \cong
S^{2}$ and $U_{1} \cap U_{2} = U_{12} \cong S^{1}$ is the equator.
More precisely, if $\vartheta$ and $\varphi$ are two angles
parametrizing points of the sphere, then $U_{1} =  
\{ 0 \leq \vartheta \leq \pi/2, 0 \leq \varphi < 2 \pi \}$, 
$U_{2} = \{ \pi/2 \leq \vartheta \leq \pi, 0 \leq \varphi < 2
\pi\}$. We take the generators of $K=SU(2)$ to be $Q_{i} = 
\tau_{i}/2$ ($i=1,2,3$), where $\tau_{i}$ are the Pauli 
matrices. The subgroup $H=U(1)$ is generated by $Q_{3}$ 
and the vector space ${\cal M}$ is spanned by
$Q_{1}$ and $Q_{2}$. Let us consider the decomposition of the
algebra ${\cal K}$ in the case under consideration. Denote by
$e_{\pm \alpha}$ the root vectors and by $h_{\alpha}$ the 
corresponding Cartan element of this algebra and take
\[
e_{\pm \alpha} = \tau_{\pm} = \frac{1}{2} \left(\tau_{1}\pm 
i\tau_{2}\right), \; \; \; 
h_{\alpha}= \tau_{3}.
\]

We choose local representatives $k^{(j)}(\vartheta, \varphi) \in
K=SU(2)$, $j=1,2$ of the
points of the coset space $S^{2} = SU(2)/U(1)$ as follows: 
\[
  k^{(1)}(\vartheta, \varphi) = e^{i\varphi \frac{\tau_{3}}{2}} 
  e^{i\vartheta \frac{\tau_{2}}{2}} 
e^{-i\varphi \frac{\tau_{3}}{2}} \; \; \mbox{and} \; \; 
  k^{(2)}(\vartheta, \varphi) = e^{-i\varphi \frac{\tau_{3}}{2}} 
  e^{i(\vartheta-\pi) \frac{\tau_{2}}{2}} 
  e^{i\varphi \frac{\tau_{3}}{2}}.
\]
By straightforward computation one obtains the forms
$\bar{\theta}_{\cal H}$ and $\bar{\theta}_{\cal M}$ as
\[
    \bar{\theta}^{(i)}  =  \left(k^{(i)}\right)^{-1} d k^{(i)} = 
    \bar{\theta}_{\cal H}^{(i)} + \bar{\theta}_{\cal M}^{(i)}.
\]

According to general formula (\ref{A-inv}) the invariant
gauge connection depends on the gauge group and the embedding
$\tau(H) \subset G$.  
The decomposition of the vector space ${\cal
M}$ into irreducible invariant subspaces of ${\cal H}$ 
corresponds to the following decomposition of representations:
\beq
    \underline{2} \rightarrow (2) + (-2),  \label{M-decomp}
\eeq
where in the r.h.s. we indicated the eigenvalues of $ad
(h_{\alpha})$, and the reducible representation carried by the
space ${\cal M}$ is indicated by its dimension in the l.h.s. 

Now let $E_{\alpha}$, $E_{-\alpha}$ and $H_{\alpha}$ be the root 
vectors and the Cartan element of the algebra $A_{1}$, which
appears as complexification of ${\cal G} = su(2)$. We assume
that they are given by the same combinations of the Pauli
matrices as the corresponding elements of complexified ${\cal
K}$ described above. Define the group homomorphism $\tau: 
H=U(1) \rightarrow G=SU(2)$ by the expression
\[
\tau \left( e^{i\alpha_{3}\frac{\tau_{3}}{2}} \right) = 
e^{i \alpha_{3} \kappa \frac{\tau_{3}}{2} } = 
\cos (\kappa \frac{\alpha_{3}}{2}) + 
i \tau_{3} \sin (\kappa \frac{\alpha_{3}}{2}).
\]  
It is easy to check that this definition is consistent if
$\kappa$ is integer. Thus the corresponding homomorphism of
algebras is labelled by $n \in Z$ and is given by 
$\tau_{n} (h_{\alpha}) = n H_{\alpha}$.
The three-dimensional space ${\cal G}$ of the adjoint
representation of $A_{1}$ decomposes into three 1-dimensional
irreducible invariant subspaces of $\tau_{n}({\cal H})$:
\beq
    \underline{3} \rightarrow (0) + (2n) + (-2n)  \label{G-decomp}
\eeq
(in brackets we indicate the eigenvalues of 
$ad \left(\tau ( h_{\alpha})\right)$). 
Compare decompositions (\ref{M-decomp}) and
(\ref{G-decomp}). For $n \neq \pm 1$ there are no equivalent
representations in the decompositions of ${\cal M}$ and ${\cal
G}$, and the intertwining operator $\phi: {\cal M} \rightarrow
{\cal G}$ is zero.  It also turns out to be zero for $n=0$.
In accordance with (\ref{A-inv}) for $n \neq \pm 1$ 
\beq
A^{(1)}_{n} = \frac{n}{2e} \tau_{3} 
       (1 - \cos \vartheta) d \varphi,  \; \; 
A^{(2)}_{n} =  -\frac{n}{2e} \tau_{3} 
       (1 + \cos \vartheta) d \varphi.
                           \label{A-SU2}
\eeq
If $n=1$ or $n=-1$ the results are more interesting.  
Let us consider the case $n=1$ in detail.
Comparing again (\ref{M-decomp}) and (\ref{G-decomp}) we see
that there are pairs of representations with the same
eigenvalues and, therefore, the operator $\phi$ is non-trivial.
It is determined by its action on basis elements of ${\cal M}$:
$\phi (e_{\alpha}) = f_{1} E_{\alpha}$, $\phi (e_{-\alpha}) = 
f_{2} E_{-\alpha}$, 
where $f_{1}$, $f_{2}$ are complex numbers. The fact that the
initial groups and algebras are compact implies a reality
condition \cite{KMRV} which tells that $f_{1} = f_{2}^{*}$. Thus,
the operator $\phi$ and the invariant connection are
parametrized by one complex parameter $f_{1}$ (we will suppress
its index from now on). Using the formulas above we obtain that
\beq
A^{(1)} = \frac{1}{2e} \left( \begin{array}{cc}
 (1 - \cos \vartheta) d\varphi & 
         f e^{-i\varphi} (-id\vartheta - \sin \vartheta d \varphi) \\
         f^{*}e^{i\varphi}(id\vartheta -  
       \sin \vartheta d \varphi) & -(1 - \cos \vartheta) d \varphi
     \end{array}  \right),    \label{A-SU2-1} 
\eeq
\beq
A^{(2)} = \frac{1}{2e} \left( \begin{array}{cc}
-(1+\cos \vartheta) d\varphi & 
        f e^{i\varphi} (-id\vartheta - \sin \vartheta d \varphi) \\
f^{*}e^{-i\varphi}(id\vartheta - 
       \sin \vartheta d \varphi) & (1+\cos \vartheta) d \varphi
     \end{array}  \right).  \label{A-SU2-2} 
\eeq
The curvature form $F$ is given by   
\beq
    F =  dA + \frac{ie}{2} [A,A] =
     - \frac{1}{2e} \tau_{3} \left( |f|^{2} - 1 \right) \sin 
    \vartheta  d\vartheta \wedge d\varphi,        \label{F-SU2}
\eeq
and the action for such configuration is equal to 
\beq
   {\cal S}_{inv}(f) = \frac{\pi}{2e^{2}} \frac{1}{R^{2}} 
    \left(|f|^{2} - 1 \right)^{2},     \label{S-SU2}
\eeq
where $R$ is the radius of the sphere. Due to the invariance
property any extremum of the action found withing the subspace of
invariant connections is also an extremum in the space of all
connections \cite{KMRV}. From Eq. (\ref{S-SU2}) we see that
there are two extrema: the trivial one with $f=0$ and the
non-trivial one with $|f|=1$. The trivial
extremum was found in Ref. \cite{RP} as a spontaneous
compactification solution in six-dimensional Kaluza-Klein
theory.

It turns out that potentials (\ref{A-SU2}), (\ref{A-SU2-1})
and (\ref{A-SU2-2}) are related to known non-abelian monopole
solutions (see, for example, \cite{CHM}). For $n
\neq \pm 1$ and for $n = \pm 1$ with $f=0$ these expressions 
coincide with the monopole solutions with the even monopole
charge $m = 2n$. 
In fact the solution with charge $m > 0$ can be transformed 
to the solution with charge $(-m)$ by the gauge transformation 
$A \rightarrow S^{-1} A S$ with the constant matrix $S=-i \tau_{1}$.  
All these monopoles live in the trivial
principal fibre bundle $P(S^{2},SU(2))$, which is the only bundle
with such structure group. The latter result also
follows from our discussion in Sect. 2. Indeed, in the case under
consideration ${\cal B}_{SU(2)}(S^{2}) = H^{2}(S^{2},
\pi_{1}(SU(2))) = 0$ since $SU(2)$ is simply connected. 
Thus, all the monopoles can be described by a unique function on the
whole sphere. This is indeed the case. Namely, there exist gauge
transformations, different for the northern and southern
patches, so that the tranformed potentials coincide. Let us
show this for the case $n=1$. In fact we will construct such 
transformations  for the whole family of
the invariant connections (\ref{A-SU2-1}), (\ref{A-SU2-2}). The
matrices giving these gauge transformations are
\[
V_{1} = i\left( \begin{array}{cc}
 \cos \frac{\vartheta}{2} & e^{-i\varphi} \sin \frac{\vartheta}{2}  \\
 e^{i\varphi} \sin \frac{\vartheta}{2}  & -\cos \frac{\vartheta}{2} 
     \end{array}  \right), \; \;    
V_{2} = i\left( \begin{array}{cc}
 e^{i\varphi} \cos \frac{\vartheta}{2}  & \sin \frac{\vartheta}{2}  \\
 \sin \frac{\vartheta}{2}  & e^{-i\varphi \cos \frac{\vartheta}{2} }
     \end{array}  \right).    
\]
By calculating $A^{(i)'} = V_{i}^{-1} A^{(i)} V_{i} + 
(ie)^{-1}V_{i}^{-1} d V_{i}$ for $i=1$ and $i=2$ one can easily 
check that the transformed fields are equal to
\beq
A^{(1)'} = A^{(2)'} = \frac{1}{2e} \left(\tau_{+} c_{+} + 
\tau_{-} c_{-} + \tau_{3} c_{3} \right)
\label{A-SU2-gen} 
\eeq 
with 
\bea
 c_{+} & = & c_{-}^{*} = e^{-i\varphi} 
 \left\{ \left[ -\sin \vartheta \cos \vartheta + 
 \left( f \cos ^{2} \frac{\vartheta}{2} 
 - f^{*} \sin^{2} \frac{\vartheta}{2} \right) 
  \sin \vartheta \right] d\varphi 
                                         \right. \nonumber \\
 & + & \left. i(-1 + f \cos ^{2} \frac{\vartheta}{2} + 
     f^{*}\sin^{2} \frac{\vartheta}{2} ) d\vartheta \right\} 
                                  \nonumber \\
 c_{3} & = & 
\left( 1 - \frac{f + f^{*}}{2} \right) \sin^{2}\vartheta d\varphi - 
i\frac{f-f^{*}}{2} \sin \vartheta d\vartheta.   \nonumber
\eea
Note that in general expressions (\ref{A-SU2-1}),
(\ref{A-SU2-2}) and (\ref{A-SU2-gen}) the phase of the complex
parameter $f$ can be rotated by residual gauge transformations
which form the group $U(1)$. 

For $f=0$ this formula gives the known expression for the $m=2$ 
$SU(2)$-monopole \cite{CHM}:
\beq
A^{(1)} = A^{(2)} = \frac{1}{4e} \left( \begin{array}{cc}
(1-\cos 2\vartheta) d\varphi & 
  e^{-i\varphi} (-2id\vartheta - \sin 2\vartheta 
 d \varphi) \\
 e^{i\varphi}(2id\vartheta - 
   \sin 2\vartheta d \varphi) & -(1-\cos 2\vartheta) d\varphi
     \end{array}  \right).  \label{A-SU2-0} 
\eeq
One can easily see that the forms (\ref{A-SU2-gen}) and (\ref{A-SU2-0}) 
do not have singularities on the sphere. 

For $f=f^{*}=1$ the form (\ref{A-SU2-gen}) vanishes. This
shows that this configuration, which is also an extremum of the
action, describes the trivial $SU(2)$-monopole with
$m=0$.  Note that in the original form the potentials
(\ref{A-SU2-1}) and (\ref{A-SU2-2}) do not seem to be trivial.
Of course, one can check that they are pure gauge configurations and
correspond to the flat connection.  Vanishing of the gauge field
(\ref{F-SU2}) for this value of $f$ confirms this.  

A similar situation occurs for $n=-1$. Again, there exists a
family of in\-va\-ri\-ant con\-nec\-tions pa\-ra\-met\-riz\-ed by a
complex parameter, say $h$, analogous to $f$. The action
possesses two extrema. One, with $h=0$, describes the
$SU(2)$-monopole solution with $m=-2$ and the other extremum, with
$|h|=1$, describes the trivial monopole with $m=0$. However, 
this family of invariant connections is gauge equivalent to the family 
for $n = +1$ and can be obtained from (\ref{A-SU2-1}), (\ref{A-SU2-2}) 
by the gauge transformation with the constant matrix $S=-i\tau_{1}$ 
and the identification $h=f$. 

The picture we obtained is the following. Various homomorphisms
$\tau_{n}: H \rightarrow G$ give rise to various invariant
connections describing $SU(2)$-monopole solutions. All monoples
are reproduced in this way. Moreover there is a continuous
family of invariant gauge connections, parametrized by one 
complex parameter $f$ which passes
through the points corresponding to $SU(2)$-monopoles with
$m=0$ ($n=0$), $m=2$ ($n=1$) and $m=-2$ ($n=-1$) in the space of
all connections of the theory. 
Classes of gauge equivalent invariant connections are labelled by  
values of $|f|$. Thus, $|f|=0$ corresponds to the class of the $m=2$ 
monopole (the $m=-2$ monopole is in the same class, as we already 
explained above), $|f|=1$ corresponds to the class of the 
$m=0$ monopole. 

Let us calculate now the contribution of the invariant
connections to the vacuum
expectation value of the Wilson loop functional. For this we
choose a loop $\gamma (\vartheta_{0})$ on $S^{2}$ given by 
\beq
 \gamma (\vartheta_{0}) = \{(\vartheta_{0}, \varphi'), 
  \vartheta_{0} = 
     \mbox{const}, 0 \leq \varphi' < 2\pi \}.  \label{loop}
\eeq
This loop is parallel to the equator, labelled by the angle 
$\vartheta_{0}$ and parametrized by the polar angle $\varphi'$. 
Here for definiteness we consider the case when the loop lies in
the northern hemisphere, i.e. $0 \leq \vartheta_{0} \leq \pi/2$.

We notice that the group element 
\[
U(\varphi) = {\cal P} \exp \left[ ie\int_{\eta(\varphi;\vartheta_{0})} 
A^{(1)}_{\phi} d\varphi' \right],
\]
where $\eta (\varphi; \vartheta_{0})$ is the same path as (\ref{loop})
but with $\varphi'$ changing from $0$ to $\varphi$, satisfies the
matrix equation 
\beq
\frac{dU(\varphi)}{d\varphi} = ie U(\varphi) A^{(1)}_{\varphi}
(\vartheta_{0},\varphi).    \label{U-eq}
\eeq
The holonomy $H_{\gamma (\vartheta_{0})} (A^{(1)}) = U(2\pi)$. 
It turns out that Eq. (\ref{U-eq}) can be solved and 
the traced holonomy is equal to 
\beq
T_{\gamma (\vartheta_{0})} (A^{(1)}) 
  =  - \cos \left( \pi \sqrt{ 1 + \frac{4S}{V} \left( 
  1 - \frac{S}{V} \right) \left( |f|^{2} -1 \right) } \right),  
  \label{T-SU2}
\eeq
where $S = 2 \pi R^{2} (1 - \cos \vartheta_{0})$ is the area of the
surface surrounded by the loop $\gamma (\vartheta_{0})$ and $V = 4
\pi R^{2}$ is the total area of the two-dimensional sphere of the 
radius $R$. 
Note that for $|f|=1$ $T_{\gamma (\vartheta_{0})}=1$ as it should
be for a flat connection. The expression (\ref{T-SU2}) is
invariant under the transformation $S \rightarrow (V - S)$. 

Finally we evaluate the quantity 
$  < T_{\gamma (\vartheta_{0})}>_{inv} = Z_{inv}
( \gamma (\vartheta_{0}) )/Z_{inv}(0), $  
characterizing the contribution of the invariant connections to
the vacuum expectation value of the Wilson loop functional,
where 
\beq
  Z_{inv}( \gamma (\vartheta_{0}) ) = 
  - \int df \  df^{*} e^{-\frac{2 \pi^{2}}{e^{2}V} 
  \left( |f|^{2} - 1\right)^{2} } 
  \cos \left( \pi \sqrt{ 1 + \frac{4S}{V} \left( 
1 - \frac{S}{V} \right) \left( |f|^{2} -1 \right) } \right). 
                               \label{Z-inv} 
\eeq
This integral is an analog of the functional integral
(\ref{Z-sum}). It mimics a ``path-integral quantization" of the
gauge model where the configuration space (the space of
connections) is finite-dimensional and consists of $SU(2)$ - invariant
connections. The integral takes into
account the contribution of the monopoles with $m=0,\pm 2$ and
fluctuations around them along the {\it invariant} direction. 
The action on the invariant connection in the formula 
above is given by Eq. (\ref{S-SU2}). 

Of course, the complete contribution of the invariant connections, 
given by Eqs. (\ref{A-SU2-1}), (\ref{A-SU2-2}) or Eq. 
(\ref{A-SU2-gen}) to the true functional integral (\ref{Z-sum}) 
is different because it includes also contributions of all 
fluctuations around them. In the next section we
are going to discuss which part of the true vacuum expectation
value $<T_{\gamma (\vartheta_{0})}>$ is captured by 
$<T_{\gamma (\vartheta_{0})}>_{inv}$. 

We would like to make a few remarks. In Ref. \cite{Mottola} it was 
argued that the Faddeev-Popov determinant on invariant
connections turns out to be zero, hence the contribution of such
connections to the functional integral vanishes. This issue was
analyzed in \cite{AGV} in a different context, namely the
authors considered similar ``quantization" 
for $SU(2)$-invariant connections in $(3+1)$ - dimensional
Ashtekar's gravity. They found that zeros of the
Faddeev-Popov determinant, which are responsible for the
vanishing of the determinant, are cancelled by the contribution
of the delta-functions of constraints and the path
integral measure is regular on invariant connections.  

\section{Discussion of the result}

In Sect. 2 and Sect. 3 we calculated the vacuum expectation
value of the Wilson loop functional. 
Let us analyze the quantity $E(g^{2}V,S/V) \equiv  
-\log < T_{\gamma (\vartheta_{0})}> $ which in a theory with fermions 
characterizes the potential of the interaction. Recall that the 
closed path $\gamma (\vartheta_{0})$ on $S^{2}$ was defined by Eq. 
(\ref{loop}), $S$ is the area of the region surrounded by this loop and 
its ratio to the total area of the surface is given by  
$S/V = (1 - \cos \vartheta_{0})/2$. $E(g^{2}V,S/V)$ as a function 
of $S/V$ for two values of $e^{2}V$ in 
the {\it abelian} case is given by Eq. (\ref{Tg}) and is presented
in Fig. 1. 

\vskip 0.1cm 
\begin{figure}[ht]
\epsfxsize=0.9\hsize
\epsfbox{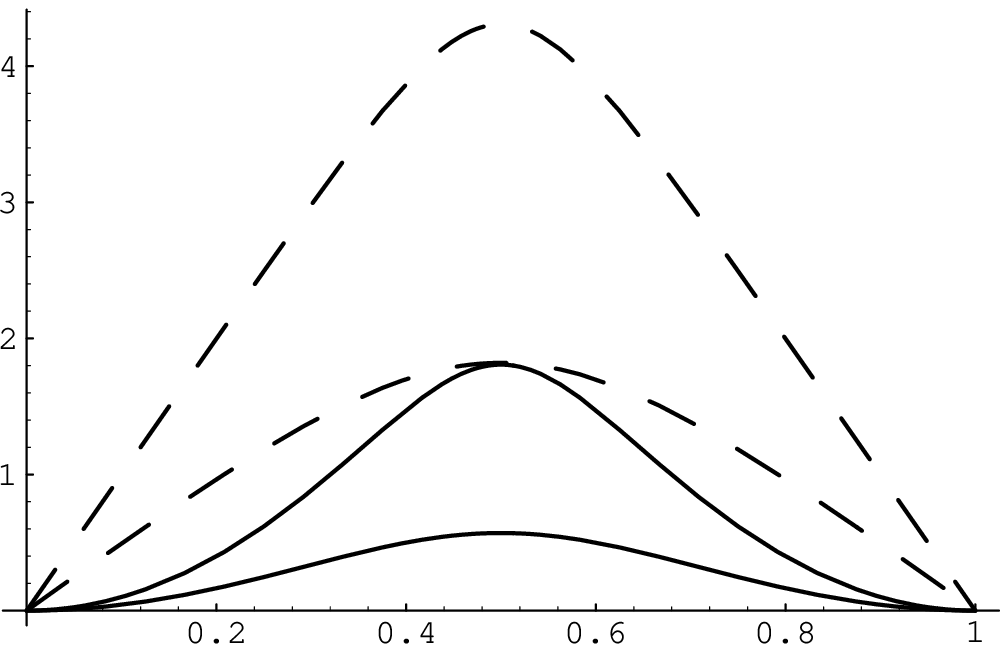}
\caption{ $E(e^{2}V,S/V)$ (dashed line) and $E_{mon}(e^{2}V,S/V)$ 
(solid line) as functions of $S/V$ for $e^{2} V=10$ (lower lines)
and $e^{2} V=20$ (upper lines) in the abelian gauge theory}
\end{figure}

In the same plot we also show the contribution
$E_{mon}(g^{2}V,S/V)$ of the {\it abelian} monopoles calculated from
(\ref{Zmon}). We see from Fig. 1 (also from the exact formula
(\ref{Zmon})) that the dependence of $E_{mon}(g^{2}V,S/V)$ on
$S/V$ at small $S/V$ is quadratic almost
till $S/V = 0.5$, where the curve reaches its maximum. 
This, being combined with the contribution of the fluctuations, 
gives the linear dependence (the area law) almost for
all $S/V$ in the interval $0 < S/V < 0.5$. 
The area law for the Wilson loop in the pure
Yang-Mills theory is considered as an indication of 
the regime of confinement in the corresponding gauge
theory with quarks. The behaviour of $E(g^{2}V,S/V)$ 
for the {\it non-abelian} theory with $G=SU(2)$
is qualitatively the same. In this case contributions of
fluctuations around a monopole depend on the monopole solution,
so that the complete $Z(\gamma)$ cannot be written in the factorized
form as in the abelian case. However, the sum of the
terms $\exp(-{\cal S}_{mon})T_{\gamma}(A_{mon})$ over all
monopoles gives rise to qualitatively the same behaviour as
$E_{mon}(g^{2}V,S/V)$ in the abelian case. 

As we have pointed out 
the linear behaviour of the total $E(g^{2}V,S/V)$ is a manifestation
of the area law for ``not very large" $S/V$, i.e. when the area
surrounded by the loop $\gamma$ is far below the half of the
total area of the compact space-time $M$. 
In this respect the quadratic behaviour of the monopole 
contributions $E_{mon}(g^{2}V,S/V)$ seems to be an important feature 
which gives rise to the linear dependence of the complete function.  

\vskip 0.1cm 
\begin{figure}[ht]
\epsfxsize=0.9\hsize
\epsfbox{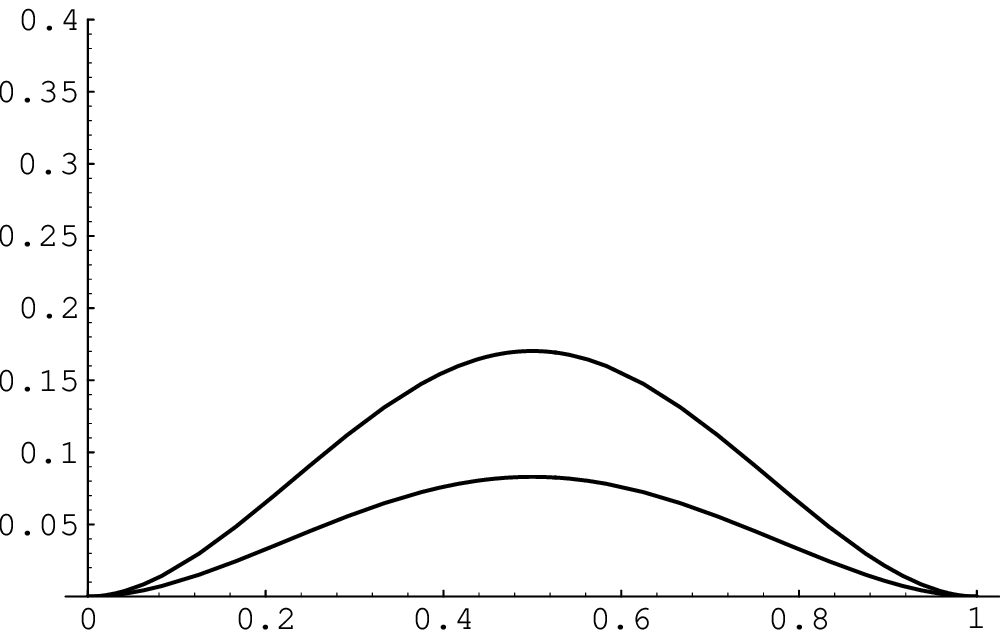}
\caption{Contribution of invariant connections $E_{inv}(e^{2}V,S/V)$ 
as a function of $S/V$ for $e^{2} V=10$ (lower line)
and $e^{2} V=20$ (upper line) in the $SU(2)$ Yang-Mills theory.}
\end{figure}

The plot of the energy $E_{inv}(g^{2}V,S/V) \equiv -\log < T_{\gamma
(\vartheta_{0})}>_{inv}$ as a function of $S/V$ is given by Eq. 
(\ref{Z-inv}) and is shown in Fig. 2. 
Comparing it with Fig. 1 we see qualitative similarity in the
contribution of invariant connections and monopoles, and 
we want to attract attention to this feature. 
It can be argued that the quadratic behaviour of the 
contribution of invariant connections, similar to that of 
the monopoles, also leads to the linear dependence of the complete 
$E(g^{2}V,S/V)$ in the $SU(2)$ case and, thus, serves as an idicator 
of the area law behaviour of Wilson loop variables in the theory.   
If this is also true in higher dimensions, one can study just the
contribution of invariant connections and from the
form of the behaviour of the function $E_{inv}$ 
get some hints about the type of the behaviour of the complete 
function $E$. We are going to study this possibility in a future
publication.
 
\bigskip

\noindent{\large \bf Acknowledgments}

We would like to thank Jose Mour\~ao and Sebasti\`a Xamb\'o
for useful discussions and valuable remarks.
J.M.A. acknowledges financial support from CICYT (project PB94-1196).  
Y.K. acknowledges financial support from 
CIRIT (Generalitat de Catalunya), RFFR (grant 96-0216413-a), 
INTAS (grant INTAS-93-1630.EXT) and JNICT 
(grant JNICT/CERN/S/FAE/1111/96).

\end{document}